\documentclass[shortbibliography,twocolumn,prl,aps,superscriptaddress,showpacs,amsmath,amssymb,floatfix]{revtex4-2}
\usepackage{bm}
\usepackage{xcolor}
\usepackage{amsfonts}
\usepackage{amsmath}
\usepackage{amssymb}
\usepackage{graphicx}
\usepackage{tabularx}
\usepackage{multirow}
\usepackage{braket}
\usepackage{commath}%
\usepackage{comment}
\usepackage[colorlinks=true, letterpaper=true, pdfstartview=FitV,  linkcolor=blue, citecolor=blue, urlcolor=blue]{hyperref}



\begin{document}

\title{Surface Originated Cross-Field Anomalous Transport in Magnetoelectric Multilayers}

\author{Jin Cao}
\affiliation{Research Laboratory for Quantum Materials, Department of Applied Physics, The Hong Kong Polytechnic University, Kowloon, Hong Kong, China}

\author{Wei Du}
\affiliation{Interdisciplinary Center for Theoretical Physics and Information Sciences (ICTPIS), Fudan University, Shanghai 200433, China}

\author{Xue-Jin Zhang}
\affiliation{Institute of Applied Physics and Materials Engineering, Faculty of Science and Technology, University of Macau, Taipa, Macau, China}

\author{Cong Xiao}
\email{congxiao@fudan.edu.cn}
\affiliation{Interdisciplinary Center for Theoretical Physics and Information Sciences (ICTPIS), Fudan University, Shanghai 200433, China}

\author{Qian Niu}
\affiliation{School of Physics, University of Science and Technology of China, Hefei, Anhui 230026, China}

\author{Shengyuan A. Yang}
\email{shengyuan.yang@polyu.edu.hk}
\affiliation{Research Laboratory for Quantum Materials, Department of Applied Physics, The Hong Kong Polytechnic University, Kowloon, Hong Kong, China}

\begin{abstract}
In material systems with slab geometry, the surface contribution to physical responses is commonly expected to diminish rapidly with increasing thickness, giving way to the bulk response.
Here, we show that this conventional wisdom is violated in a class of gate-induced responses, including gate-induced orbital and spin magnetization as well as cross-field anomalous thermoelectric transport.
We develop a general framework for these effects, which naturally decomposes the total response into surface- and bulk-contributions treated on equal footing. Remarkably, the volume-averaged surface contribution remains finite in the thick-slab limit and exhibits the same thickness scaling as the bulk term. Furthermore, the surface response originates from band geometric quantities distinct from those in the bulk, being constrained solely by surface symmetries. As a result, it can dominate the overall response when the bulk contribution is symmetry-forbidden. Taking MnBi$_2$Te$_4$ multilayers as an example, we predict a strong surface-dominated cross-field anomalous Nernst effect arising from surface Berry curvature, which is readily accessible to experimental detection.
These findings reveal a previously overlooked significance of surface response and open a new direction in the study of surface quantum geometry.
\end{abstract}

\maketitle

For condensed-matter transport experiments, a common setup is to fabricate the material of interest into a thin slab geometry and sandwich it with dual electric gates. The advance in nano-fabrication technology has enabled precise control of sample thickness and atomically clean surfaces/interfaces, especially for the class of van der Waals layered materials~\cite{Burch2018Magnetism,Mak2019Probing,Kurebayashi2022Magnetism}.
Meanwhile, the dual gating technique enables at least two functions. One is to tune the carrier density, and the other is to apply an electric field in the vertical direction. Although this gate field is normal to the transport plane, it may induce symmetry breaking, trigger unusual transport phenomena, and even drive phase transitions~\cite{Ohno2000Electric,Pan2015Electric,Deng2018Gate,Wang2018Electric,Burch2018Electric,Jiang2018Electric,Huang2018Electrical,Collins2018Electric,Verzhbitskiy2020Controlling,Du2020Berry,Yu2020Valley,Gao2021Layer,Liu2024OME,Yu2025EHE}.

In general, a physical effect of a finite system involves contributions from both bulk and boundaries. For a slab geometry which is confined in the vertical direction, the boundary contribution is mostly associated with the top and bottom surfaces. For ultrathin samples (with only a few atomic layers), it is difficult and not very meaningful to distinguish bulk from surfaces. With increasing sample thickness $L$ (usually  for $L>10$ or 20 atomic layers), such distinction becomes important, and since the local environments at bulk  and at surface are different, their contributions to response effects may exhibit different behaviors. Indeed, several recent experimental studies reported interesting transport signals that are attributed to surface contributions of thin layers~\cite{He2021quantum,Kumar2021Room,Lee2024,Yang2025ultrabroadband}. Nevertheless, if further increasing thickness $L$, the common wisdom is that the bulk contribution should eventually overwhelm the surface contribution~\cite{Ong2011TI,Syers2015Kondo}. This is intuitive from a scaling argument: The bulk contribution to transport is $\propto L$, whereas the surface contribution $\propto L^0$. In other words, if normalized by thickness, the response coefficient should be $\propto L^0$ for bulk, yet $\propto L^{-1}$ for surface.
Thus, the surface contribution is commonly believed to be negligible for thick samples.

In this work, we discover that a class of gate-field-induced response phenomena actually \emph{violate} this common wisdom.
Examples include the gate-induced orbital magnetization and cross-field anomalous thermoelectric transport effects. Our key findings contain the following. (i) We propose a new unified route to decompose a response coefficient into two parts, dubbed as layer-polarized (LP) and layer-hybridized (LH) terms. The significance of this dichotomy is that the two parts quickly
acquire the meaning of surface and bulk contributions as the system thickness increases, allowing the surface and bulk responses to be treated on equal footing. (ii) Remarkably, for the discussed effects, we show that the surface and bulk contributions possess the same scaling with $L$. Specifically, the surface response coefficient scales as $L^0$, which cannot be neglected compared to bulk contribution even in the $L\rightarrow \infty$ limit. (iii) We clarify the symmetry conditions for the studied effects, especially the possibility that the bulk contribution is forbidden and the response is solely contributed by the surface. (iv) The surface contributions are determined by quantum geometric quantities distinct from the bulk responses. For example, for cross-field anomalous transport, the surface response originates from surface Berry curvature, whereas the bulk response is determined by the bulk Berry-connection polarizability~\cite{Gao2014,Liu2022}.

These points are demonstrated in a concrete system, the MnBi$_2$Te$_4$ multilayers. We predict strong surface-dominated cross-field transport signals that can be verified in experiment. Our work provides a unified framework for studying surface and bulk responses, uncovers previously unexpected significance of surface contribution
that challenges the common wisdom, reveals distinct quantum geometric origins of surface and bulk responses, and proposes effects that can be used to probe the surface quantum geometry.

\begin{figure}
\begin{centering}
\includegraphics[width=6.2cm]{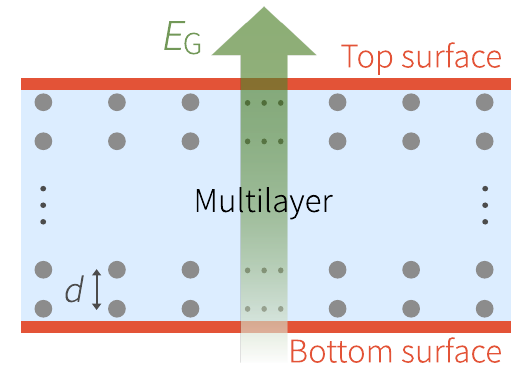}
\par\end{centering}
\caption{\label{Fig_1} A multilayer system under gate field $E_G$. The layer here may correspond to either real atomic sites or a lattice chosen for discretizing the low-energy model of the material system. }
\end{figure}

\emph{\textcolor{blue}{LP and LH responses.}}
Consider a multilayer material system with a slab geometry, i.e., being extended along $x$ and $y$ directions and confined along $z$ direction, having a thickness of $L$. Suppose the electronic Hamiltonian $H_0$ of the system in the absence of applied fields is known. Such a model can always be realized as a tight-binding model on a lattice, as illustrated in Fig.~\ref{Fig_1}. Here, each layer of the lattice may physically correspond, e.g., to active atomic sites of a layered material; but more generally, they may just be an artificial lattice chosen for discretizing the given model $H_0$.

Under a gate electric field $\bm E_G=E_G\hat{z}$, it adds a term
$\Delta H_G=- E_G \hat{\varrho}$
to the lattice model. Here, $\hat{\varrho}\equiv -ed\mathcal J$ is the layer polarization operator, where $-e$ is the electron charge, $d$ is the lattice constant along $z$, and $\mathcal J$ is the layer-pseudospin matrix \cite{zheng2025layer}: For a system having $N$ layers, $\mathcal J=\text{diag}\left( \frac{N - 1}{2}, \frac{N -
3}{2}, \ldots, - \frac{N - 3}{2}, - \frac{N - 1}{2} \right)$. In the following, we consider weak $E_G$ (i.e., not causing electric breakdown) and focus on effects linear in $E_G$.

\begin{figure}
\begin{centering}
\includegraphics[width=7.5cm]{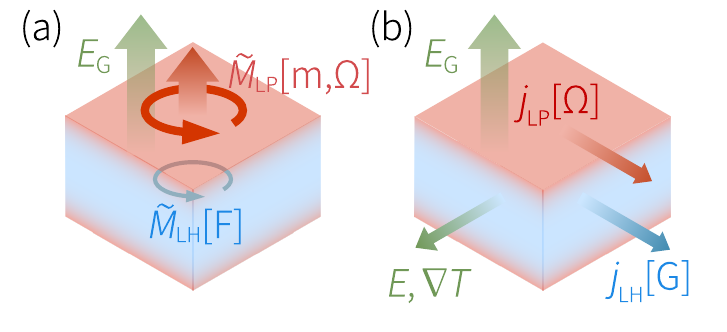}
\par\end{centering}
\caption{\label{Fig_2}Schematics of (a) gate-field-induced orbital magnetization $\tilde{M}$ and (b) cross-field anomalous thermoelectric transport. The responses consist of surface (LP) and bulk (LH) contributions. The symbols in the square brackets highlight their dependence on distinct band geometric quantities.}
\end{figure}

As the first example, we consider the effect of gate-induced electronic orbital magnetization (Fig.~\ref{Fig_2}(a)), described by the magnetoelectric coupling coefficient
\begin{eqnarray}\label{chi}
  \chi \equiv \frac{\partial \tilde M}{\partial E_G}\Big|_{E_G=0},
\end{eqnarray}
where $\tilde M$ is the orbital magnetization in the vertical direction in the presence of $E_G$.
It is known that $\tilde M$ consists of two terms~\cite{Xiao2006,Xiao2010Berry}:
$
  \tilde M=\tilde M^m+\tilde M^\text{BC}.
$
$\tilde M^m$ arises from the orbital magnetic moment of Bloch electrons
\begin{eqnarray}\label{Mm}
  \tilde M^m=\int [d\bm k]f_{n\bm k} \tilde m_{n}(\bm k),
\end{eqnarray}
where $[d\boldsymbol{k}]\equiv\sum_{n}d^2\boldsymbol{k}/(2\pi)^{2}$ is a shorthand notation, $\bm k$ is the 2D wave vector for the multilayer system, $n$ is the band index, $f$ is the distribution function, and $\tilde m_n(\bm k)$ is the orbital magnetic moment of a Bloch wave packet centered at state $|n\bm k\rangle$.
$\tilde M^\text{BC}$ is a Berry curvature contribution:
\begin{eqnarray}\label{MBC}
  \tilde M^\text{BC}=-\frac{e}{\hbar}\int [d\bm k]g_{n\bm k} \tilde \Omega_{n}(\bm k),
\end{eqnarray}
where $g=-k_B T\ln [1+\exp ((\tilde\varepsilon-\mu)/k_B T)]$ is the grand potential density, $\mu$ is chemical potential, and $\tilde \Omega$ is the Berry curvature. The functions $f$ and $g$ depend on the band energy that is corrected by gate field \cite{fan2024dipole,Yu2025EHE}:
$\tilde \varepsilon_n(\bm k)=\varepsilon_n-E_G\varrho_n$,
where $\varepsilon_n$ is the unperturbed energy, and the correction is proportional to the \emph{diagonal} elements $\varrho_n$ of the layer polarization matrix
$
  \varrho_{n\ell}\equiv\langle n\bm k|\hat{\varrho}|\ell\bm k\rangle
$.
For simple notations, we do not write out $k$-dependence explicitly in following equations, where quantities are understood to be evaluated at the same $k$ point.

The tildes in $\tilde m$ and $\tilde \Omega$ indicate these quantities are also corrected by gate field. To linear order in $E_G$, we find (details in Supplemental Material~\cite{supp})
\begin{equation}
\tilde m_{n}(\bm k)=m_{n}-e(\boldsymbol{v}_{n}\times\boldsymbol{a}_{n}^{E_{G}})_{z}+E_{G}\mathcal{F}_{n}.\label{orbi_moment}
\end{equation}
Here, $m$ is the unperturbed value in the absence of $E_G$, the other two terms are the first-order correction. In the second term, $\bm v$ is the band velocity, $\boldsymbol{a}^{E_{G}}=E_G \bm{\mathcal{G}}$ is the gate-induced Berry connection, with
\begin{equation}
\bm{\mathcal{G}}_{n}(\bm k)=2\hbar\ \mathrm{Im}\sum_{\ell\neq n}\frac{\varrho_{n\ell}\boldsymbol{v}_{\ell n}}{\left(\varepsilon_{n}-\varepsilon_{\ell}\right)^{2}}
\label{Berry curvature}
\end{equation}
being the Berry-connection polarizability~\cite{Gao2014,Liu2022} defined for a multilayer system. In the last term,
\begin{equation}
\mathcal{F}_n(\bm k)=-2\mathrm{Re}\sum_{\ell\neq n}\frac{\varrho_{n\ell}m_{\ell n}}{\varepsilon_{n}-\varepsilon_{\ell}}-\frac{e}{2\hbar}\left(\nabla_{\boldsymbol{k}}\times\bm{\mathcal{Q}}_{n}\right)_{z},\label{AOP}
\end{equation}
where $m_{\ell n}=e\sum_{n'\neq n}[(\boldsymbol{v}_{\ell n'}+\delta_{\ell n'}\boldsymbol{v}_{n})\times\boldsymbol{\mathcal{A}}_{n' n}]_{z}/2$ represents the interband orbital magnetic moment, with $\boldsymbol{\mathcal{A}}_{n' n}$ being the interband Berry connection, and
$
\bm{\mathcal{Q}}_{n}=\mathrm{Re}\sum_{\ell\neq n}\varrho_{n\ell}\boldsymbol{\mathcal{A}}_{\ell n}$.
We note that under a vertical $B$ field, the electron acquires an out-of-plane electric dipole moment of $B\mathcal F$, so
$\mathcal{F}$ has the meaning of anomalous orbital polarizability~\cite{Gao2014,Gao2015Geometrical,Xiao2021Adiabatic,Wang2024IPHE,Wang2024orbital} for a multilayer system.
Meanwhile, the gate-field-corrected Berry curvature is given by
\begin{eqnarray}\label{BC}
  \tilde \Omega_{n}(\bm k)=\Omega_n + (\nabla_{\bm k} \times {\bm a}^{E_{G}}_n)_z,
\end{eqnarray}
where the correction can also be expressed in terms of $\bm{\mathcal{G}}$. Interestingly, the corrections of the geometric quantities $\tilde m$ and $\tilde \Omega$ involve only the \emph{off-diagonal} elements $\varrho_{n\ell}$ with $n\neq\ell$.

Inserting results (\ref{Mm}-\ref{BC}) into $\tilde M$, one obtains the expression of $\chi$.
As mentioned above, each of these term depends on {either} the diagonal {or} the off-diagonal elements of $\hat \varrho$ operator, so it is natural to classify terms into two parts: $\chi=\chi^\text{LP}+\chi^\text{LH}$, with
\begin{eqnarray}
\chi^{\text{LP}}&=&\frac{\partial}{\partial \mu}\int [d\boldsymbol{k}] \varrho_n \mathcal M_n,\label{LP}\\
\chi^{\text{LH}}&=&\int [d\boldsymbol{k}] f_{0}\mathcal{F}_n,
\label{LH}
\end{eqnarray}
where $\mathcal M_n=f_0 m_n-\frac{e}{\hbar}g_0 \Omega_n$, and subscript $0$ in $f_0$ and $g_0$ means they depend on the unperturbed energy $\varepsilon_n$.

The LP response $\chi^{\text{LP}}$ depends on diagonal elements $\varrho_n$, so it is enhanced by states with a large layer polarization. In comparison, the LH response involves only off-diagonal elements $\varrho_{n\ell}$, which require states distributed across multiple layers, i.e., layer-hybridized states. This explains their naming, and suggests that the two parts should exhibit distinct behaviors, as we shall see below.

\emph{\textcolor{blue}{Surface vs bulk contributions.}} To unveil the characters of LP and LH responses, we consider systems with large thickness $L$. In this limit, one can distinguish bulk and surface of the system. In the interior of the bulk, the energy eigenstates are fully layer hybridized, i.e., having equal distribution across layers, which leads to $\varrho_n=0$. This shows the bulk states contributes to $\chi^\text{LH}$ but not $\chi^\text{LP}$. On the other hand, large $\varrho_n$ requires layer polarized states, which can only be associated at top and bottom surfaces.
This argument demonstrates that for thick systems, $\chi^\text{LP}$ and $\chi^\text{LH}$ just correspond to the surface and bulk contributions, respectively. As $\chi^\text{LP}$ and $\chi^\text{LH}$ are unambiguously defined based on a clear dichotomy, this offers us a unified way to study surface and bulk responses on equal footing.

Moreover, this treatment allows an easy demonstration of the unconventional scaling of surface response.
As mentioned, usually, the surface contribution should scale as $L^{-1}$ (when normalized by thickness), hence becomes negligible for large $L$. Is this the same for the current case? The answer is No. Here, the surface response, $\chi^\text{LP}$, is associated with the top/bottom surface, for which $\varrho_n\sim \mp eL/2$.
Thus, the normalized surface response in three-dimensional (3D) units
\begin{eqnarray}
  \chi^{\text{LP}}_{\mathrm{3D}}\equiv\chi^{\text{LP}}/L \propto L^{0},
\end{eqnarray} i.e., it has the same $L^0$-scaling behavior as bulk contribution, and it does not decay with thickness.
This is a remarkable feature, showing that for such a response, the surface contribution {cannot be neglected} even for thick samples.

In the $L\rightarrow \infty$ limit, the surface contribution simplifies to
\begin{eqnarray}
    \chi^{\text{LP}}_{\mathrm{3D}}=-\frac{e}{2} \frac{\partial}{\partial \mu} (M_t-M_b),\label{LP:surface-OM}\\
  M_{t/b}=\int [d\boldsymbol{k}] (f_0 m_{t/b}-\frac{e}{\hbar}g_0\Omega_{t/b}),
\end{eqnarray}
where $m_{t/b}$ and $\Omega_{t/b}$ denote the orbital moment and Berry curvature for top/bottom surface states. In real materials, the approach towards the limiting value in Eq.~(\ref{LP:surface-OM}) as $L$ increases is actually very rapid, which will be demonstrated in a while.

\emph{\textcolor{blue}{Cross-field anomalous thermoelectric transport.}}
Our proposed classification (into LP and LH) and identified unconventional scaling behavior are general for the class of
gate-field-induced responses. Here, we consider the examples of cross-field anomalous transport effects, i.e., a transverse electric current $\bm j$ that is bilinear in the longitudinal driving $E$ field (or temperature gradient $\nabla T$) and in the gate field (see Fig.~\ref{Fig_2}(b)):
\begin{equation}
    \bm j= \alpha_0 \bm E_{G} \times \bm E -\alpha_1 \bm E_{G} \times \nabla T,
\end{equation}
where $\alpha_i$'s ($i=0,1$) are the response coefficients. And we shall focus on the intrinsic responses, which represent inherent properties of the multilayer.
These effects are intimately connected to the gate-driven magnetoelectricity discussed above, because they can be viewed as the
anomalous Hall (Nernst) transport accompanying the magnetization induced by $E_G$.
In fact, $\alpha_i$'s can be directly obtained from the Berry curvature part of magnetization~\cite{Xiao2006,Xiao2020unified}. For the current case, we find
\begin{eqnarray}\label{16}
     \alpha_0 =e\frac{\partial^2 \Tilde{M}^\text{BC}}{\partial E_{G} \partial \mu}, \quad
    \alpha_1 =-\frac{\partial^2 \Tilde{M}^\text{BC}}{\partial E_{G} \partial T}.
\end{eqnarray}

We can again decompose the response coefficients into LP and LH parts:
$\alpha_i =\alpha_i^{\text{LP}}+\alpha_i^{\text{LH}}$. Using Eqs.~(\ref{MBC}) and (\ref{16}), we obtain compact formulas
\begin{eqnarray}
\alpha^{\text{LP}}_i & =&-\frac{e^2}{\hbar}\int [d\boldsymbol{k}]\varrho_{n}\Omega_{n}\left(-\frac{\varepsilon_n-\mu}{eT}\right)^i f_{0}^{\prime}, \label{LPi}\\
\alpha^{\text{LH}}_i & =&-e^2\int[d\boldsymbol{k}]\left(\boldsymbol{v}_n\times\bm{\mathcal{G}}_n\right)_{z}\left(-\frac{\varepsilon_n-\mu}{eT}\right)^i f_{0}^{\prime}.
\label{Nernst}
\end{eqnarray}
In line with the foregoing discussion, for thick systems, LP and LH parts correspond respectively to surface and bulk responses. Indeed, in the $L\rightarrow\infty$ limit, if we replace the off-diagonal $\varrho_{n\ell}$ with the interband Berry connection $-e(\mathcal{A}_{z})_{n\ell}$, then the normalized LH response $\alpha^{\mathrm{LH}}_{0,\mathrm{3D}}\equiv \alpha^{\mathrm{LH}}_0/L$ will recover exactly the intrinsic nonlinear Hall conductivity of a bulk material~\cite{Gao2014,Wang2021,Liu2021Intrinsic,Xiao2025definition}. This correspondence can be proved formally, as shown in the Supplemental Material \cite{supp}.

Similarly, the surface responses $\alpha_i^\text{LP}$ exhibit the same $L$-scaling as bulk and cannot be neglected. Normalized by thickness,
\begin{eqnarray}
  \alpha^{\mathrm{LP}}_{i,\mathrm{3D}}\equiv \alpha_i^{\mathrm{LP}}/L\propto L^{0}
\end{eqnarray}
is independent of $L$ in the $L\rightarrow\infty$ limit. For example
\begin{equation}
    \alpha^{\text{LP}}_{0,\mathrm{3D}}=\frac{e^3}{2\hbar}\int [d\boldsymbol{k}] \left(\Omega_t - \Omega_b \right)f_0',
    \label{AHE topological}
\end{equation}
which is determined by the Berry curvature of surface-state Fermi surfaces.
Particularly, if the surface states have a more or less isotropic Fermi surface, the result can be simplified as ($i=0,1$)
\begin{eqnarray}
    \alpha^{\text{LP}}_{i,\mathrm{3D}}=\frac{e^3}{2\hbar} \left(\frac{\pi^2k^2_BT}{3e} \frac{\partial}{\partial \mu}\right)^i\left(D_t\Omega_t - D_b\Omega_b \right),
    \label{AHE Dirac surface}
\end{eqnarray}
where $D$ denotes the density of states. In (\ref{AHE Dirac surface}), the surface Nernst conductivity $\alpha^{\text{LP}}_{1,\mathrm{3D}}$ is obtained by using Sommerfeld expansion.

\begin{figure}[t]
\begin{centering}
\includegraphics[width=8.5cm]{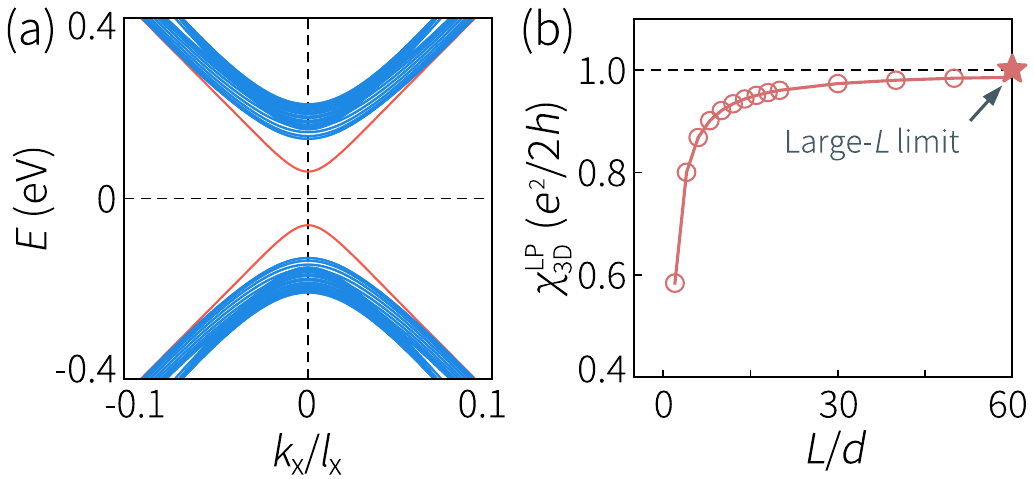}
\par\end{centering}
\caption{\label{Fig_3}(a) Band structure of MnBi$_2$Te$_4$ multilayer model with a thickness of 20 septuple layers. The red color highlights the surface bands. (b) Calculated $\chi^{\text{LP}}_{\mathrm{3D}}\equiv\chi^{\text{LP}}/L$ (at $\mu=0$) as a function of the slab thickness $L$ (taking even-number of layers). The star indicates the value in the $L\rightarrow\infty$ limit. The solid curve serves as a guide to the eyes.}
\end{figure}

\emph{\textcolor{blue}{Scenarios with dominant surface response.}}
The gate-field-induced effects discussed above, namely, the magnetoelectricity and cross-field anomalous transport,
share the same symmetry character. Their response coefficients are time-reversal-odd pseudoscalars. This requires the multilayer system having broken time-reversal ($\mathcal T$) symmetry, i.e., it must be magnetic. Furthermore, its point group should (i) \emph{not} contain any improper rotations (ii) nor any symmetry $\mathcal T R$, with $R$ being a proper rotation. Groups satisfying (i) are known as chiral groups. In view of this, symmetry groups obeying both (i) and (ii) may be termed as magnetochiral groups. Out of the 122 magnetic point groups, there are 32 such magnetochiral groups, as listed in Supplemental Material~\cite{supp}.

We highlighted the significance of surface contribution in these effects.
It is interesting to have a scenario, where the bulk contribution is somehow suppressed, so the surface  dominates the response. For the cross-field transport, this happens when the bulk is insulating, while the surface is metallic.
Another possibility is when the bulk response is forbidden by some symmetry, while the surface is not (due to its reduced symmetry). This is the case for MnBi$_2$Te$_4$ multilayers.

\emph{\textcolor{blue}{Application to MnBi$_2$Te$_4$ multilayers.}} Consider even-layer MnBi$_2$Te$_4$~\cite{Gong2019,Xu2019,Otrokov2019,Otrokov2019Unique,Liu2020Robust,Ye2021,Chang2021MBT,Gao2021Layer}, which has an A-type antiferromagnetic ordering. In its bulk, i.e., assuming periodicity in all three directions, it has
$\mathcal{T}\tau_{1/2}$ symmetry ($\tau_{1/2}$ is half lattice translation along $z$ axis), so its magnetic point group contains $\mathcal T$, not belonging to a magnetochiral group. In contrast, this  $\mathcal{T}\tau_{1/2}$ symmetry is evidently broken at top and bottom surfaces. Therefore, regarding the effects studied here,
the bulk (LH) contribution is symmetry-forbidden, and the responses must be dominated by surface (LP) contribution.

Figure~\ref{Fig_3} shows the results for gate-driven orbital magnetization. In the calculation, we adopt the widely employed tight-binding model for even-layer MnBi$_2$Te$_4$~\cite{Zhang2019Topological,supp}. In Fig.~\ref{Fig_3}(a), we show the low-energy band structure for a 20-layer  MnBi$_2$Te$_4$, in which one can distinguish bulk states (blue color) and surface states (red color). Figure~\ref{Fig_3}(b) shows the plot of  $\chi^{\text{LP}}_{\mathrm{3D}}(\equiv\chi^{\text{LP}}/L)$ as a function of $L$, at intrinsic Fermi level $\mu=0$. One observes that with increasing $L$,
$\chi^{\text{LP}}_{\mathrm{3D}}$ quickly approaches an asymptotic value of $e^2/2h$, which recovers the well-known bulk-limit value for axion
insulators~\cite{Qi2008Topological,Essin2009Magnetoelectric,Qi2011Topological,Zhang2019Topological,Vanderbilt2018Berry} (which can also be exactly obtained from our formula (\ref{LP:surface-OM})).

\begin{figure}
\begin{centering}
\includegraphics[width=8.5cm]{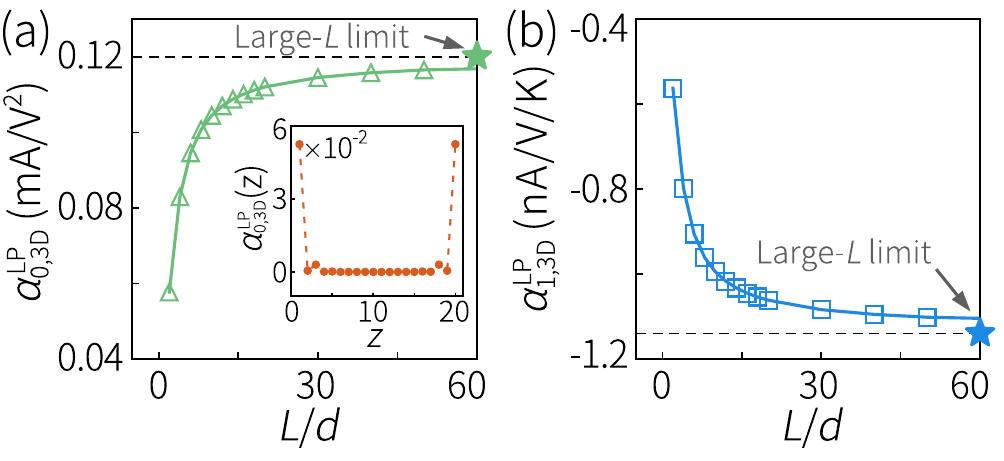}
\par\end{centering}
\caption{\label{Fig_4}(a) Calculated cross-field anomalous Hall coefficient $\alpha^{\mathrm{LP}}_{0,\mathrm{3D}}$ for MnBi$_2$Te$_4$ multilayers as a function of $L$ (even-number of layers), at $\mu=0.1~\mathrm{eV}$. The inset shows the layer-resolved contribution to $\alpha^{\mathrm{LP}}_{0,\mathrm{3D}}$ for the case with $L=20$. The star indicates the value for $L\rightarrow\infty$ limit. (b) Same as (a), but for the Nernst response $\alpha^{\mathrm{LP}}_{1,\mathrm{3D}}$.}
\end{figure}

Next, consider the cross-field anomalous transport. In Fig.~\ref{Fig_4}, we plot the transport coefficients $\alpha^{\mathrm{LP}}_{i,\mathrm{3D}}(\equiv \alpha_i^{\mathrm{LP}}/L)$ versus $L$. Similar to $\chi$, they quickly saturate towards the bulk-limit values, as indicated by the solid stars calculated from Eq.~(\ref{AHE Dirac surface}).
Actually, this surface originated response is remarkably strong. For example, at thickness larger than ten layers, $\alpha^{\text{LP}}_{1,\mathrm{3D}}\sim -1~\mathrm{nA/(V\cdot K)}$ at $T=15~$K. Under moderate $E_G$ of 0.1~V/nm, the anomalous Nernst conductivity $\alpha_1 E_{G}$ can reach $-0.1$~A/(K$\cdot$m), which is comparable to the strongest low-temperature anomalous Nernst effects reported so far~\cite{ANE2018Co2MnGa,ANE2019CoSnS-Felser,ANE2017Mn3Sn-Behnia,ANE2017Mn3Sn}.
At the same time, the $E_G$ induced magnetization ($\sim 0.1$~mT) is small, compared to previously reported systems \cite{ANE2018Co2MnGa,ANE2019CoSnS-Felser,ANE2017Mn3Sn-Behnia,ANE2017Mn3Sn}. Such a strong Nernst effect realized with weak net magnetization (hence weak stray field) is a desired feature for device applications.

\emph{\textcolor{blue}{Discussion.}}
We have unveiled that surface response plays a particularly significant role in a class of gate-field-induced effects in multilayer systems: it exhibits unconventional scaling behavior and can completely surpass the bulk response.
This contrasts with the common wisdom that bulk effects are typically dominant. The dominant surface responses identified here open a new avenue for probing surface quantum geometric properties, such as the surface Berry curvature.
It is also a great advantage that such responses are highly tunable by gating and by surface/interface engineering. As illustrated in the case of MnBi$_2$Te$_4$, the magnitude of the cross-field anomalous Nernst effect can be strong.
Taken together, these findings highlight a promising direction for future research.

We have developed a general framework
for classifying and analyzing surface and bulk contributions in multilayers, which can be readily applied to other gate induced phenomena. For example, beyond charge transport, the analysis can be directly extended to anomalous heat transport by electrons:
\begin{eqnarray}
  \bm j^h/T=\beta_1\bm E_G\times \bm E-\beta_2\bm E_G\times\nabla T,
\end{eqnarray}
where $\bm j^h$ is the heat current density. The coefficients $\beta_i$'s can again be divided into LP and LH parts. In fact, it is easy to show that $\beta_1=\alpha_1$ and $\beta_2=\alpha_2$ (with $i=2$ in Eqs.~(\ref{LPi},\ref{Nernst})).
In addition, in discussing the gate driven magnetoelectricity, we focused on the orbital magnetization, whereas the spin magnetization can also be treated using our framework. A difference from the orbital magnetoelectricity is that for spin, its surface (LP) response
requires a Fermi surface, and must vanish if surface states are insulating. Thus, for systems like multilayer MnBi$_2$Te$_4$, its gate-driven magnetoelectricity is purely from surface orbital response if the surface is insulating.
In Supplemental Material~\cite{supp}, we present an estimation of the surface spin responses for the case of metallic surface of MnBi$_2$Te$_4$ multilayers.

It is worth noting that in the effects discussed here, the surface and bulk responses differ in a fundamental way, because they are related to fundamentally different band geometric properties. This is distinct from most other response effects, e.g., the longitudinal electric transport. This feature also distinguishes our finding from previously reported nonlinear surface transport effects~\cite{He2021quantum,Kumar2021Room,Lee2024,Yang2025ultrabroadband,Sodemann2021arc,Manchon2022arc}, which actually share the same quantum geometric origin as bulk response. 

\emph{\textcolor{blue}{Data availability.}}
The data that support the findings of this article are not publicly available. The data are available from the authors upon reasonable request.

\end{document}